\documentclass[twocolumn,aps,prl,superscriptaddress,showpacs,amsmath,amssymb,floats]{revtex4}

\usepackage{amsmath}
\usepackage{amssymb}
\usepackage{graphicx}
\usepackage{keyval}
\usepackage{dcolumn}
\usepackage{bm}
\usepackage{color}
\usepackage{txfonts}
\usepackage[]{sidecap}

\begin{document}

\title{Direct observation of the bandwidth control Mott transition in  the NiS$_{2-x}$Se$_x$ multiband system}

\author{H. C. Xu}
\author{Y. Zhang}
\author{M. Xu}
\author{R. Peng}

\affiliation{State Key Laboratory of Surface Physics, Department of Physics, and
Advanced Materials Laboratory, Fudan University, Shanghai 200433,
People's Republic of China}

\author{V. N. Strocov}
\author{M. Shi}
\author{M. Kobayashi}
\affiliation{Swiss Light Source, Paul Scherrer Institute, CH-5232 Villigen PSI, Switzerland}

\author{B. P. Xie}
\email{bpxie@fudan.edu.cn}

\author{D. L. Feng}
\email{dlfeng@fudan.edu.cn}

\affiliation{State Key Laboratory of Surface Physics, Department of Physics, and
Advanced Materials Laboratory, Fudan University, Shanghai 200433,
People's Republic of China}

\begin{abstract}

The bulk electronic structure of NiS$_{2-x}$Se$_x$  is studied  across the bandwidth-control Mott transition (BCMT) by soft X-ray angle-resolved photoemission spectroscopy.  The microscopic picture of the  BCMT in this multiband non-half-filled system is revealed for the first time.  We show that  Se doping does not alter the Fermi surface volume. When approaching the insulating phase with decreasing Se concentration, we observed that the Fermi velocity continuously decreases. Meanwhile, the coherent quasiparticle weight continuously decreases and is transferred  to higher binding energies, until it suddenly disappears across the Mott transition. In the insulating phase, there is still finite spectral weight at the Fermi energy, but it is incoherent and dispersionless due to strong correlations. Our results provide a direct observation of BCMT, and unveil its distinct characters in a multiband non-half-filled system.

\end{abstract}


\maketitle

Mott transition is a metal-insulator transition (MIT) due to strong electron correlations. When the ratio between the  on-site Coulomb repulsion U and the bandwidth W is larger than a certain critical value, the electron hopping between neighboring sites is prohibited \cite{Review}.   As demonstrated in the single-band Hubbard model,  large U splits a half-filled band  into a filled lower Hubbard band (LHB) and an empty upper Hubbard band (UHB).
Furthermore, as  illustrated by dynamical mean-field theory (DMFT) calculations  for intermediate U \cite{Zhang93},  the metallic phase exhibits a complex spectral function with Hubbard bands at high energies and quasiparticle bands at low energies. The Mott transition occurs when  the quasiparticle mass diverges and its  spectral weight   is transferred completely to the Hubbard bands with increasing U or decreasing W, as suggested earlier by Brinkman and Rice  \cite{BrinkmanRice}.

One could realize  Mott transition  by controlling either the filling or the bandwidth of the Hubbard bands  \cite{Review}. In filling control, partially-filled Hubbard band is created by doping, as in the cuprate high temperature superconductors. In  bandwidth control, W is tuned  against U by applying physical or chemical pressure, as inferred indirectly from the transport properties of V$_2$O$_3$, NiS$_{2-x}$Se$_x$, etc. \cite{VO,Honig1998RV}.
The spectral weight transfer and band evolution predicted by DMFT have been observed in Ca$_{1-x}$Sr$_x$VO$_3$  \cite{CSVO,teipeiPRB}, however, since it  is metallic over
the entire doping range, MIT is never reached.
So far, there is no direct observation of the  electronic structure evolution across the bandwidth-control Mott transition (BCMT).

\begin{figure}[t]
\includegraphics[width=8.7cm]{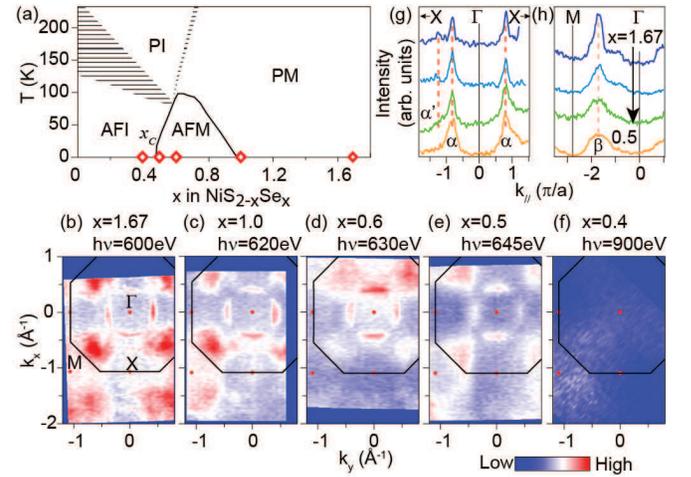}
\caption{(color online). (a) The  phase diagram of NiS$_{2-x}$Se$_x$ reproduced from Ref.~\onlinecite{Matsuura2000PD}. The   measured dopings are marked by the diamonds.  AFI, AFM, PI, and PM stand for antiferromagnetic insulator, antiferromagnetic metal, paramagnetic insulator, and paramagnetic metal, respectively. (b)-(f)  The Fermi surface mapping for various dopings in the $\Gamma$-X-M plane. The intensities are integrated over [$E_F$-15~meV, $E_F$+15~meV] and then normalized by the integrated intensity at -4~eV. (g),(h) The MDC's at $E_F$ normalized by spectrum intensities for various dopings along $\Gamma$-X (g), and  $\Gamma$-M (h), respectively.
}
\label{DDFS}
\end{figure}

NiS$_{2-x}$Se$_x$ is a prototypical BCMT system, with the MIT at $x_c=0.43$ \cite{NiSSePure,XPS,Matsuura2000PD} (Fig.~\ref{DDFS}(a)). Due to its non-half-filled multiband nature, its BCMT is more sophisticated than those described by the  half-filled single-band Hubbard model, and a consistent and comprehensive understanding has not been achieved.
According to the previous transport measurements, the approach to the antiferromagnetic insulator (AFI)  phase with doping is characterized by an effective mass enhancement in the paramagnetic metal (PM) regime (Fig.~\ref{DDFS}(a)), and a decrease in carrier density in the antiferromagnetic metal (AFM) regime \cite{JPTrans}. The decrease in carrier density was attributed to the shrinkage of Fermi surface by antiferromagnetic order \cite{JPTrans}. However, the Fermi surface  has never been resolved, let alone its doping evolution, leaving this scenario unverified.
On the other hand, in previous ultraviolet angle-resolved photoemission (UV-ARPES) studies, a narrow band of many-body nature develops at the Fermi energy ($E_F$) in metallic phase\cite{Matsuura1996ARPES}. However, this narrow band persists in the insulating regime, which was attributed to possible carrier doping by Se/S vacancies \cite{Matsuura1998TBHARPES}, or   some  surface states,  since UV-ARPES is surface sensitive  \cite{Sarma2003Surface}.  Recently,  the bonding-antibonding splitting of S-S (Se-Se) dimer bands was suggested determining  the gap, and a novel scenario of $p$-gap driven MIT was proposed against  the BCMT  \cite{Kunes2010p_gap}. The electronic structure evolution of  this prototype  system across its MIT thus needs a thorough examination.

\begin{figure}[bt]
\includegraphics[width=8.7cm]{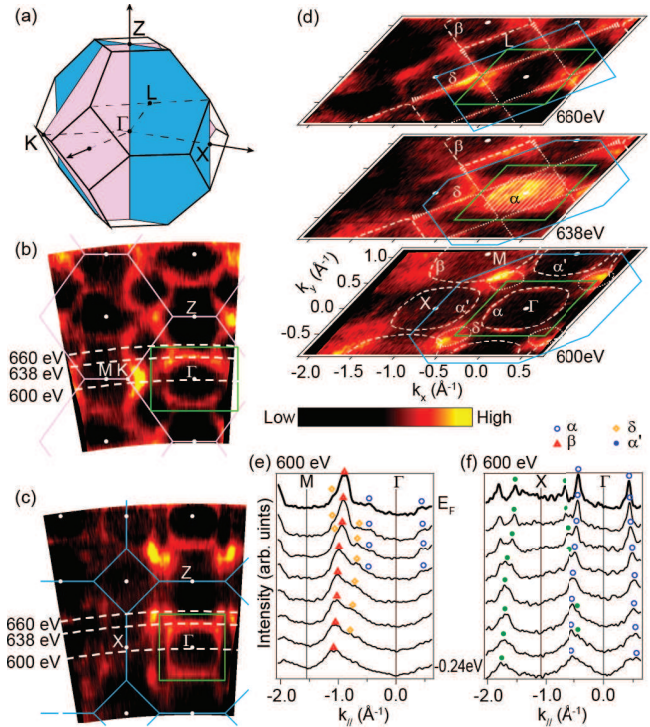}
\caption{(color online). (a) The Brillouin zone (BZ) for the face-centered-cubic sublattice of Ni$^{2+}$. (b),(c) The Fermi surface mapping in the $\Gamma$-M-Z plane, and the $\Gamma$-X-Z plane, respectively.  The green rectangles illustrate the 1st BZ for the simple cubic lattice.  (d) Fermi surface slices in the $\Gamma$-X-M plane ($h\nu$=600~eV), the plane at the half BZ height (660~eV), and a intermediate plane (638~eV), respectively, as indicated by dashed curves in (b) and (c). All these intensities were integrated over [$E_F$-15~meV, $E_F$+15~meV]. The Fermi surface sheets are shown by dashed curves. Part of the $\delta$ Fermi surface is hard to trace, as marked by the thicker dashed curves. The hatched area indicates the Fermi patch formed by cutting along the Fermi surface.
(e),(f) Momentum distribution curves (MDC's) near $E_F$ along $\Gamma$-M (e), and $\Gamma$-X (f), respectively. }
\label{FS}
\end{figure}

We have investigated the bulk electronic structure of NiS$_{2-x}$Se$_x$ by soft X-ray angle-resolved photoemission spectroscopy (SX-ARPES). The SX-ARPES provides much better $k_z$ resolution and bulk sensitivity as its main advantages over the UV-ARPES \cite{Strocov,Strocov_VSe2_ADDRESS,Razzoli}, thus suitable for studying the bulk electronic structure of NiS$_{2-x}$Se$_x$, where strong three-dimensional characters are expected due to its cubic structure. The Fermi surfaces are resolved for the first time with a doping-independent Luttinger volume in the metallic regime.  When approaching the AFI phase, we observed that the bandwidths and the Fermi velocities ($V_F$'s) of the coherent quasiparticle bands decrease and eventually diminish across the MIT. Meanwhile, the coherent spectral weight is transferred into higher binding energies, and eventually depleted entirely at the MIT in contrast to the surface sensitive UV-ARPES results. In addition to directly prove the BCMT in NiS$_{2-x}$Se$_x$, we also reveal the distinct electronic signature of BCMT in a non-half-filled multiband system.

Single crystals of NiS$_{2-x}$Se$_x$ were synthesized, with dopings covering all phases in the phase diagram (Fig.~\ref{DDFS}(a)) \cite{Matsuura2000PD}. Those with x$>$0.71 were grown by chemical vapor transport method \cite{Vapor}, and the others by Te flux method \cite{Flux}. The photoemission data were taken with $p$-polarized light at the Advanced Resonant Spectroscopies (ADRESS) beam line of the Swiss Light Source (SLS) \cite{Strocov_ADDRESS}. The overall energy resolution is about 80~meV.  Samples were cleaved \textit{in situ} at the (001) plane and measured at 11~K under a vacuum better than $5\times 10^{-11}~mbar$.

\begin{figure*}[t]
\includegraphics[width=17.5cm]{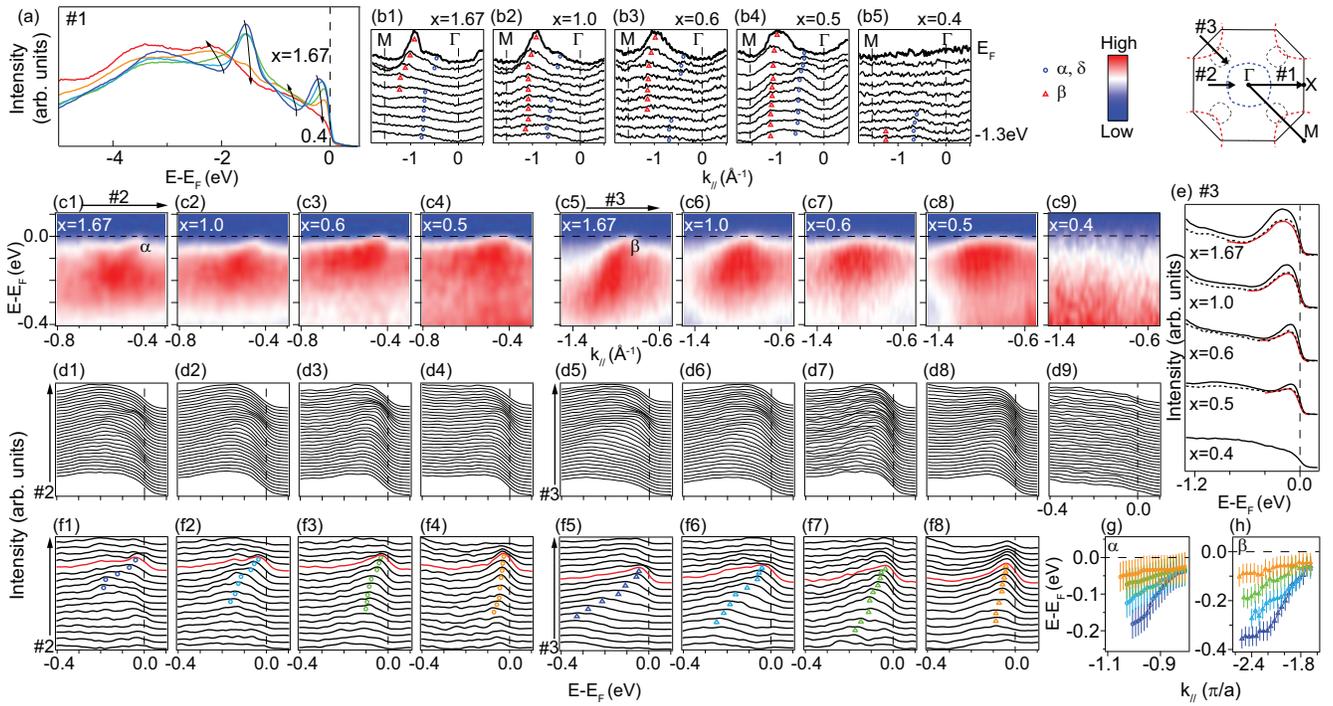}
\caption{(color online).  (a) Doping dependence of the spectra integrated along M-$\Gamma$-X (cut \#1 in the inset) normalized by total intensity. (b1)-(b5) MDC's along $\Gamma$-M for doping x=1.67, 1.0, 0.6, 0.5, and 0.4, respectively. The blue open circle and red triangle marks stand for features corresponding to the bands $\alpha$ or $\delta$, and  $\beta$, respectively. (c1)-(c4) Intensity along cut \#2 as indicated in the inset for doping x=1.67, 1.0, 0.6, and 0.5, respectively. (c5)-(c9) Intensity along cut \#3 for doping x=1.67, 1.0, 0.6, 0.5, and 0.4, respectively. (d1)-(d9) Energy distribution curves (EDC's) corresponding to (c1)-(c9). (e) Integrated EDC's along cut \#3 (black solid curves) plotted with EDC's at $\Gamma$ (black dashed curves), and the incoherent/dispersionless background (red solid curves). (f1)-(f8) The EDC's of (c1)-(c8) after subtracting the  incoherent background in panel (e). The  open circles and  triangles trace the dispersions of $\alpha$ and $\beta$, respectively. (g),(h)  The extracted dispersions, as color-coded in (f1)-(f8), for  $\alpha$ (g) and  $\beta$ (h). The inset illustrates the cuts \#1, \#2 and \#3 in the $\Gamma$-X-M plane.
}
\label{reNorm}
\end{figure*}

\begin{figure}[t]
\includegraphics[width=8.7cm]{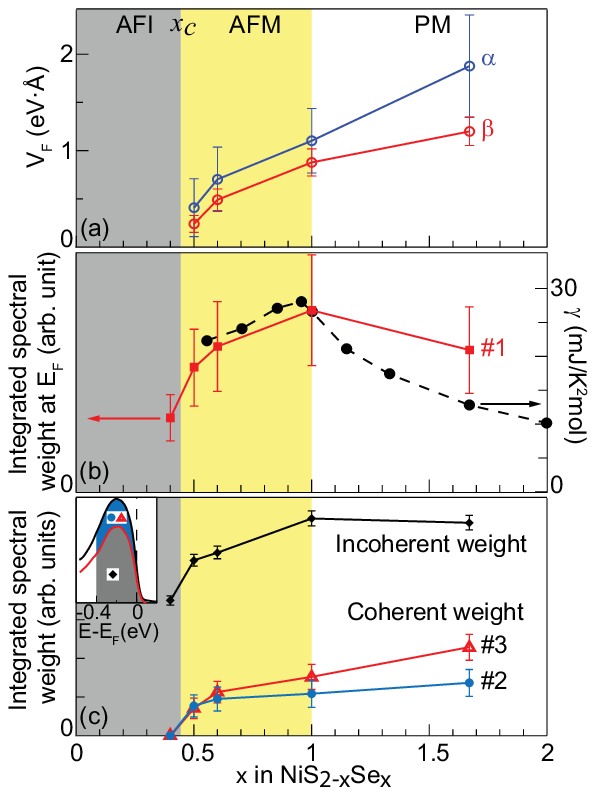}
\caption{(color online). (a) The doping dependence of Fermi velocities ($V_F$) of band $\alpha$ and band $\beta$ (Figs.~\ref{reNorm}(f1)-\ref{reNorm}(f8)). (b) The doping dependence of spectral weight at $E_F$ integrated over [$E_F$-40~meV, $E_F$+40~meV]  along path \#1 in Fig.~\ref{reNorm}, which resembles the specific heat linear coefficient $\gamma$ \cite{JPTrans}.
 (c) The doping dependence of the incoherent/dispersionless background (gray area illustrated in the inset) and the dispersive spectral weight along cuts \#2 and \#3 (blue area illustrated in the inset) over energy range [$E_F$-0.4~eV, $E_F$+0.1~eV]. }
\label{Summary}
\end{figure}

The Brillouin zone (BZ) corresponding to the face-centered-cubic (fcc) sublattice of Ni$^{2+}$ in NiS$_{2-x}$Se$_x$ is shown in Fig.~\ref{FS}(a) \cite{Supple}. Along the two high symmetric $\Gamma$-M-Z (pink) and $\Gamma$-X-Z (blue) planes (Fig.~\ref{FS}(a)), $k_z$ dependent photoemission intensity distribution of NiS$_{0.33}$Se$_{1.67}$ at $E_F$ are shown in Figs.~\ref{FS}(b) and \ref{FS}(c), with the finite photon momentum corrected. After assuming the inner potential ($V_0$) to be 25~eV, the periodicity of the Fermi surface  matches that of  the BZ for the fcc sublattice, rather than that of the BZ for simple cubic lattice \cite{Supple}. Three cross-sections of the Fermi surface taken with different photon energies in Fig~\ref{FS}(d) show their three-dimensional structure. The  dispersions near $E_F$ are illustrated  by the momentum distribution curves (MDC's) along $\Gamma$-M and $\Gamma$-X  in Figs.~\ref{FS}(e) and \ref{FS}(f). We find that the Fermi surface in the $\Gamma$-X-M plane (lower panel of Fig.~\ref{FS}(d)) are composed  of three sets: 1. a hole pocket centered at $\Gamma$ contributed by the band $\alpha$, and a hole pocket centered at X noted as  $\alpha'$ that is the counterpart of $\alpha$ in the 2nd simple cubic BZ; 2. an electron pocket contributed by  the band $\beta$ centered at M; 3. four small and weak electron pockets at the corner of simple cubic BZ, contributed by the band $\delta$. The observed complex electronic structure contradicts the recent local-density approximation calculations, which give just one electron pocket around $\Gamma$ in NiSe$_2$ \cite{Kunes2010p_gap}. Hereafter, we focus our discussion on the data in the $\Gamma$-X-M plane, which represent the general electronic structure.

Noting the lattice constant varies with doping from  5.96~{\AA}  (NiSe$_2$) to  5.69~{\AA}  (NiS$_2$)  \cite{Vegard}, we pinpoint the photon energy that corresponds to the $\Gamma$-X-M plane for each doping from high-$k_z$-resolution photoemission intensity distributions. The Fermi surface mapping in the $\Gamma$-X-M plane for all measured dopings are shown in Figs.~\ref{DDFS}(b)-\ref{DDFS}(f), and  there is no detectable variation of Fermi surface in metallic regime, except for the fading of $\alpha$' pocket possibly due to the enhanced dimer disorder. The Fermi crossings along $\Gamma$-X (Fig.~\ref{DDFS}(g)) and $\Gamma$-M (Fig.~\ref{DDFS}(h)) appear invariant. On the other hand, the intensity at $E_F$ decreases gradually with decreasing $x$, and the Fermi surfaces disappear in the insulating phase. The doping independent Fermi surface volume proves the iso-valent nature of the S/Se substitution, which also suggests  the validity of the Luttinger theorem with progressively increasing correlations.  Therefore, the decrease in carrier density is not due to the the proposed Luttinger volume reduction \cite{JPTrans,Iwaya2004STM}, but the diminishing spectral weight at $E_F$.

In Fig.~\ref{reNorm}(a), the valence band spectra integrated over cut \#1 show two narrow peaks at -1.6~eV and just below $E_F$, respectively. They are composed of both Ni $3d$ and S/Se~$p$ states according to our resonant photoemission data, confirming the charge transfer nature of the band structure \cite{Supple,Guillot1977RSNi,XPS,Fujimori1996RSPE}. With decreasing $x$, one could  observe a reduction of  the peak near $E_F$,  accompanied by a relative enhancement of spectral weight at 0.5-1~eV below $E_F$. The  peak at -1.6~eV behaves almost identically, whose  spectral weight  seems to be transferred to higher binding energies as well, suggesting enhanced inelastic scattering with decreased doping.
Unlike the previous UV-ARPES studies \cite{Matsuura1996ARPES,Matsuura1998TBHARPES}, the peak  near $E_F$ disappears in
our SX-ARPES data for x=0.4, the AFI phase. Considering the surface sensitivity of UV-ARPES, the persistent narrow peak observed there may be largely originated from the surface \cite{Matsuura1998TBHARPES}.

In the common picture of BCMT, bandwidth is the key parameter, whose evolution can only be extracted from the bare band dispersion.
Nevertheless, as illustrated before both theoretically \cite{Mishchenko_TJ,Rosch_Bareband} and experimentally for the blue bronze and cuprate \cite{Grioni_TSe2,Yang_Bi2201,Xie2007Waterfall}, the bare band dispersion is manifested as the MDC's centroid of the broad incoherent spectrum.
Indeed, the MDC's  in Figs.~\ref{reNorm}(b1)-\ref{reNorm}(b5)  show highly dispersive features with rather weak doping dependence  for NiS$_{2-x}$Se$_x$, thus presumably the bare bandwidth is not strongly varied by S substitution of Se. This is consistent with the recent band calculations that the bandwidth decreases  by  less than 7\%  from   NiSe$_2$ to NiS$_2$  \cite{Kunes2010p_gap}.
On the other hand, in the first 0.4~eV below $E_F$, one could observe the doping dependence of the spectral intensity distribution along cut \#2 (Figs.~\ref{reNorm}(c1)-\ref{reNorm}(c4)) and \#3 (Figs.~\ref{reNorm}(c5)-\ref{reNorm}(c9)). The corresponding energy distribution curves are shown in Figs.~\ref{reNorm}(d1)-\ref{reNorm}(d9), respectively, where
the spectra of the metallic samples are made of a large incoherent/dispersionless broad background and a weak dispersive feature, while that of the insulating sample is featureless.

To obtain the incoherent/dispersionless background, one could either take the broad spectrum at $\Gamma$ (since one would not expect any feature near $E_F$ there), or combine the momentum-independent part at individual binding energies \cite{Supple}. The similar results from these two methods are plotted in Fig.~\ref{reNorm}(e), together with the angle-integrated spectra along cut \#3 for comparison.
After subtracting the background,  one can readily identify that the Fermi velocities of the $\alpha$  and $\beta$ bands  decrease with decreasing doping  in Figs.~\ref{reNorm}(f1)-\ref{reNorm}(f8), as summarized in Figs.~\ref{reNorm}(g)-(h).
We note that the dispersionless feature  shifts towards $E_F$ with decreasing $x$ (Fig.~\ref{reNorm}(e)),  following the   renormalization of the coherent bands, which implies that the incoherent weight could be some shake-off side bands from the coherent quasiparticles by strong  correlations.

The Fermi velocities of the quasiparticle dispersion along both cut \#2 and \#3 are summarized in Fig.~\ref{Summary}(a) as a function of doping. They drop rapidly when approaching the Mott transition, and most likely diminish at the transition based on the trend presented here.  Such a strong enhancement or even diverging quasiparticle mass near $x_c$ is consistent with previous transport measurements \cite{JPTrans}.
Therefore, even though the bare band cannot be extracted accurately,  the observed Fermi velocity or quasiparticle bandwidth renormalization provides a straightforward evidence of BCMT \cite{Review}. Particularly, the rate of band renormalization is much faster than the bare bandwidth variation, highlighting the dominant role of the increased strong correlations.

In Fig.~\ref{Summary}(b), we plot the density of state represented by the integrated spectral weight at $E_F$ along cut \#1 in Fig.~\ref{reNorm}. As expected, it shows similar doping dependence as the specific heat linear coefficient $\gamma$ reproduced from Ref.~\onlinecite{JPTrans}. Both quantities  peak at $x=1$, the critical point between the PM and AFM phase, where the spin fluctuations lead to large $\gamma$ \cite{JPTrans}.  The density of state at $E_F$ does not exhibit any anomaly  across the Mott transition, and  it is still finite even for the insulator. Similarly in Fig.~\ref{Summary}(c),  the incoherent spectral weight (integrated over the window as illustrated in the inset) dominates the spectrum at lower dopings, and crosses the Mott transition smoothly.  In contrast, the coherent weight diminishes at Mott transition for both cut \#2 and \#3. Intriguingly, the decreasing rate are accelerated near the Mott transition for both the coherent and incoherent weight.

It is clear that NiS$_{2-x}$Se$_x$ can be described neither  by a simple band insulator picture, nor  by the traditional single-band Hubbard model with the splitting of a half-filled band and a clean gap.
Instead, the BCMT of this non-half-filled multiband system could be viewed as a coherent-incoherent transition, where the coherent  weight is governed by the Mott physics as described in the picture of Brinkman and Rice \cite{BrinkmanRice}, while the rest incoherent spectral weight gradually dominates at $E_F$ due to rapidly increasing strong correlations. In  this way,  the system is converted from a bad metal to a bad insulator, as found by resistivity and optical measurements \cite{Kautz1972,Yao1996PD}.
Similar coherent-incoherent MIT  characterized by finite but incoherent spectral weight at $E_F$ in the insulating phase has been observed before in polaronic systems \cite{Kwizera1980polaron,Valla2002coh-incoh,zhangyanFeTePRB}.

To summarize, we have presented the first direct demonstration and main electronic characters of the BCMT in a multiband non-half-filled system, NiS$_{2-x}$Se$_x$.
We found that  the Fermi surface volume is unaffected by sulfer doping in the PM and AFM phase, and the bare bandwidth is just moderately narrowed. The increased  correlations  transfer
the spectral weight into  higher binding energies, reduce the coherent bandwidth,  suppress the coherent spectral weight, and eventually lead to the Mott transition that is characterized by  a divergent effective quasiparticle mass and a depleted coherent weight at $E_F$. Moreover, the  insulating phase  is characterized by  finite incoherent spectral weight  at $E_F$ without opening a  charge gap.  These results deepen our understanding of  Mott transition in general.

We gratefully acknowledge Prof. Hide Takagi for providing some single crystals to us years ago at the beginning of this project, and Prof. Zhengyu Weng and Prof. Qianghua Wang for helpful discussions. This work is supported in part by the National Science Foundation of China and National Basic
Research Program of China (973 Program) under the grant Nos. 2012CB921400,
2011CB921802, 2011CBA00112.

\end{document}